\documentclass[
notoc,nohyper]{JHEP3} 

\newcommand{\Gammait}{{\mit\Gamma}}

\newcommand{\tr}{\mathop{\rm tr}\nolimits}

\newcommand{\SU}{\mathop{\rm SU}}
\newcommand{\SO}{\mathop{\rm SO}}
\newcommand{\U}{\mathop{\rm {}U}}
\newcommand{\Real}{\mathop{\rm Re}\nolimits}
\newcommand{\rmd}{{\rm d}}

\newcommand{\ring}{\mathaccent"7017 }

\newcommand\fverb{\setbox\pippobox=\hbox\bgroup\verb}
\newcommand\fverbdo{\egroup\medskip\noindent%
                        \fbox{\unhbox\pippobox}\ }
\newcommand\fverbit{\egroup\item[\fbox{\unhbox\pippobox}]}
\newbox\pippobox

\title{
Two dimensional $\mathcal{N}=(2,2)$ super Yang-Mills theory on the lattice
via dimensional reduction}

\author{Hiroshi Suzuki\\
Institute of Applied Beam Science, Ibaraki University, Mito 310-8512,
Japan\\
Theoretical Physics Laboratory, RIKEN, Wako 2-1, Saitama 351-0198,
Japan\thanks{Address after October 1st, 2005.}\\
E-mail: \email{hsuzuki@riken.jp}}

\author{Yusuke Taniguchi\\
Institute of Physics, University of Tsukuba, Tsukuba, Ibaraki 305-8571,
Japan\\
E-mail: \email{tanigchi@het.ph.tsukuba.ac.jp}}

\received{\today}               
\accepted{\today}               

\preprint{IU-MSTP/70\\RIKEN-TH-48\\
UTHEP-507\\UTCCS-P-14}

\abstract{The $\mathcal{N}=(2,2)$ extended super Yang-Mills theory in
2~dimensions is formulated on the lattice as a dimensional reduction of a
4~dimensional lattice gauge theory. We use the plaquette action for a
bosonic sector and the Wilson- or the overlap-Dirac operator for a fermion
sector. The fermion determinant is real and, moreover, when the
overlap-Dirac operator is used, semi-positive definite. The flat directions
in the target theory become compact and present no subtlety for a numerical
integration along these directions. Any exact supersymmetry does not exist
in our lattice formulation; nevertheless we argue that one-loop calculable
and finite mass counter terms ensure a supersymmetric continuum limit to all
orders of perturbation theory.}


\begin{document} 

\maketitle 

\section{Introduction}
In this paper, we consider a lattice formulation of the $\mathcal{N}=(2,2)$
super Yang-Mills (SYM) theory in 2~dimensions,\footnote{This theory is
sometimes referred to as the $\mathcal{N}=2$ SYM theory in 2~dimensions,
though this usage of terminology is somewhat confusing.} which is one of
subjects of recent
developments~\cite{Kaplan:2002wv,Sugino:2003yb,Catterall:2004np} (see also
refs.~\cite{Giedt:2003ve}--\cite{Onogi:2005cz} for recent related
works).\footnote{In the present paper, the gauge group~$G$ is taken to be
$\SU(N_c)$.} In these works, ingenious constructions, based on the
orbifolding and deconstruction~\cite{Kaplan:2002wv}, topological field
theoretical representations~\cite{Sugino:2003yb} and the twisted
supersymmetry and a geometrical discretization~\cite{Catterall:2004np},
respectively, are applied to find lattice actions that are invariant under a
nilpotent supersymmetry.\footnote{For an application of the twisted
supersymmetry to lattice supersymmetric theories from a somewhat different
viewpoint, see ref.~\cite{D'Adda:2004jb}.} Then it is found that, at least in
2~dimensional extended SYM theories, invariance under a full set of
supersymmetry is restored in the continuum limit without any tuning of
parameters.\footnote{This general strategy for supersymmetric theories on
the lattice was advocated in ref.~\cite{Catterall:2001fr}. For the preceding
works with a similar strategy, see ref.~\cite{Elitzur:1982vh}. Analyses in
refs.~\cite{Kaplan:2002wv,Sugino:2003yb} show that 3~dimensional extended
SYM theories generically require a tuning of parameters.}

However, if one considers a numerical implementation of those constructions
in refs.~\cite{Kaplan:2002wv,Sugino:2003yb,Catterall:2004np}, a fact that
the fermion determinant in these formulations is not guaranteed to be real
(see ref.~\cite{Giedt:2003ve}, for example) may pose a serious problem. In
this paper, from a quite different viewpoint, we propose yet another lattice
formulation of the $\mathcal{N}=(2,2)$ SYM theory in 2~dimensions which is
free from this complex determinant problem. By doing this, we aim at a
rather practical (if not theoretically intriguing) lattice formulation of
this system.

Our basic idea was inspired by a work of Fujikawa~\cite{Fujikawa:2002ic} and
proceeds as follows: The spacetime lattice provides an ultraviolet (UV)
cutoff for correlation functions. In perturbation theory with this lattice
regularization, an integrand of a Feynman integral associated to a Feynman
diagram is modified by the lattice spacing~$a$ so that the integral is UV
convergent. Now suppose that we have a Feynman diagram in the continuum
theory whose associated Feynman integral is UV finite. For such an integral,
we may remove the lattice cutoff as $a\to0$ because the integral must be
independent of any UV cutoff when it is sent to infinity. This argument
indicates that, in the $a\to0$ limit, only potentially UV diverging Feynman
integrals and correlation functions are influenced by details of a lattice
formulation.

Our present target theory, the $\mathcal{N}=(2,2)$ SYM theory in
2~dimensions, is perturbatively super-renormalizable. According to the
standard power counting, besides vacuum bubble diagrams, only one-loop one-
and two-point functions of bosonic fields are potentially UV diverging.
One-point functions (tadpoles) are forbidden by the gauge invariance which
will be manifest in our lattice formulation. Hence, only one-loop two-point
functions of bosonic fields, which are potentially logarithmically
diverging, may be influenced by a lattice regularization. In a power series
expansion of these two-point functions with respect to the external
momentum, only the first constant term is logarithmically diverging and the
rest are UV finite. The gauge invariance again forbids this first term for
gauge fields. We thus expect that only mass terms of scalar fields are
influenced by a lattice formulation. In other words, in the $a\to0$ limit,
any trail of a lattice formulation, in particular a breaking of the
supersymmetry in our present problem, will be eliminated by tuning a
coefficient of scalar mass terms. Moreover, this tuning will be calculable
in {\it one-loop\/} perturbation theory.\footnote{A similar idea for lower
dimensional SYM theories can be found in the discussion of
ref.~\cite{Golterman:1988ta}.}

Lattice artifacts are of $O(a)$ and scalar two-point functions diverge as
$O(\log a)$ at most. Thus we expect that any trail of a lattice formulation
in the $a\to0$ limit, even if it exists, is of~$O(a^0)$. In summary, we
expect that an addition of mass counter terms of scalar fields, whose
coefficient is calculable in the one-loop order and UV finite, ensures a
supersymmetric continuum limit. No further tuning of parameters will be
required.

It turns out that the above argument based on the continuum power counting
is a little bit too naive. If one makes an ``exotic'' choice of a lattice
action, there can appear certain diagrams, being peculiar to lattice
perturbation theory, that are not covered by the continuum power
counting.\footnote{For example, to a lattice action of a 2~dimensional
$g\phi^3$ theory, one may add an ``irrelevant" interaction such as
$\sqrt{g}a\phi\partial_\mu\phi\partial_\mu\phi$ which induces a tadpole of
$O(\sqrt{g}/a)$ and two point function of $O(ga^0)$; these spoil the
continuum loop expansion and the continuum power counting.} Therefore, after
fixing a definite form of our lattice action in later sections, we confirm
that such an ``exotic'' does not occur with our lattice action, by using the
Reisz power counting theorem~\cite{Reisz:1987da} for lattice Feynman
integrals.

Accepting temporarily the above argument based on the continuum power
counting, it is clear that a lattice action for the $\mathcal{N}=(2,2)$ SYM
theory in 2~dimensions (in our scenario) is large extent arbitrary. (A
coefficient of mass counter terms of course depends on a lattice action
chosen.) By using this wide freedom, we can avoid the above problem of
complex fermion determinant. For definiteness, we start with a lattice
formulation of the $\mathcal{N}=1$ SYM theory in 4~dimensions, in which the
plaquette action and the Wilson- or the overlap-Dirac
operator~\cite{Neuberger:1998fp} are used. Then by a dimensional reduction,
we obtain a lattice action for the $\mathcal{N}=(2,2)$ SYM theory in
2~dimensions. In this construction, the fermion determinant, even with
presence of gauge and scalar fields, is real and, moreover, when the
overlap-Dirac operator is used, semi-positive definite.\footnote{The
domain-wall fermion~\cite{Kaplan:1992bt} with an infinite 5~dimensional
extent shares this feature and may be adopted in our formulation as well. We
will not compute a corresponding coefficient of mass counter terms however.}
Another bonus of this construction is that scalar fields in the
2~dimensional theory, which are originally gauge fields along the reduced
dimensions, are compact.\footnote{It is interesting to note that if all
bosonic fields including scalar fields are compact in a lattice formulation
with an {\it exact\/} nilpotent supersymmetry, then the Neuberger no go
theorem~\cite{Neuberger:1986xz} on a lattice BRS symmetry would imply a
vanishing Witten index~\cite{Witten:1982df}. H.S. would like to thank
Fumihiko Sugino for reminding this point.} Thus there is no subtlety
associated with an integration along the flat directions for which the
classical potential energy vanishes. A non-compactness of scalar fields in
the target theory is restored in the continuum limit.\footnote{There is
however a subtle issue whether lattice formulations based on non-compact
scalar fields and compact scalar fields are in the same universality class;
we do not consider this issue in the present paper.} These features of our
formulation must be desirable for practical numerical simulations.

The above argument crucially depends on perturbation theory and we ignored
a possible subtlety associated with the infrared (IR) divergences in this
2~dimensional massless theory. For the first point, we have no further
comment and simply assume a validity of perturbation theory in a weak
coupling phase. Note that in the present model, a dimensionless coupling
constant $ag_0$ goes to zero in the continuum limit. On the second point, a
careful treatment of zero modes will show that our program proceeds as
expected.

This paper is organized as follows: In section~2, we summarize basic facts
concerning the dimensional reduction and the $\mathcal{N}=(2,2)$ SYM theory
in 2~dimensions, in particular on the one-loop effective potential in a
finite box. In section~3, our lattice construction is presented. By
comparing the one-loop effective potential in our lattice framework with
that of the continuum theory, we determine a coefficient of required scalar
mass counter terms. It is found that the coefficient is IR as well as UV
finite. In section~4, we discuss further prospects and possible
generalizations.

\section{Target continuum theory}
\subsection{Dimensional reduction and the $\mathcal{N}=(2,2)$ SYM in
2~dimensions}
The $\mathcal{N}=1$ SYM theory takes a particularly simple form when the
spacetime dimension~$d$ is 4, 6 and~10~\cite{Brink:1976bc}. For example,
the on-shell fields of the $\mathcal{N}=1$ SYM theory in 4~dimensions
consist of the gauge boson and the adjoint Majorana fermion. Moreover, by
applying the dimensional reduction~\cite{Scherk:1975fm} to this theory in
$d=4$, the classical action of the $\mathcal{N}=2$ SYM theory in $d=3$, the
$\mathcal{N}=(2,2)$ SYM theory in $d=2$ and the $\mathcal{N}=4$ ``SYM''
theory in $d=1$ is obtained~\cite{Brink:1976bc}.

A similar statement holds in a spacetime with the euclidean signature,
although in the 4~dimensional euclidean space the Majorana condition cannot
be imposed in an $\SO(4)$ invariant way. The action of the $\mathcal{N}=1$
$\SU(N_c)$ SYM theory in $d=4$ euclidean space would be\footnote{Our
convention: Anti-hermitian generators~$T^a$ of $\SU(N_c)$ are normalized as
$\tr\{T^aT^b\}=-(1/2)\delta_{ab}$. The totally anti-symmetric structure
constants are defined by $[T^a,T^b]=f_{abc}T^c$ and
$f_{abc}f_{abd}=N_c\delta_{cd}$ for $\SU(N_c)$. The capital Roman indices,
$M$, $N$, \dots\ run over 0, 1, 2 and~3, and the Greek indices, $\mu$, $\nu$,
\dots\ run over 0 and~1. A summation over repeated indices is understood
unless noted otherwise. Dirac matrices for $d=4$ (denoted by~$\Gamma_M$) and
for $d=2$ (denoted by~$\gamma_\mu$) are hermitian and obey
$\{\Gamma_M,\Gamma_N\}=2\delta_{MN}$
and~$\{\gamma_\mu,\gamma_\nu\}=2\delta_{\mu\nu}$. The chiral matrices are
defined by $\Gamma=-\Gamma_0\Gamma_1\Gamma_2\Gamma_3$
and~$\gamma=-i\gamma_0\gamma_1$. $\hat M$ and $\hat\mu$ denote a unit vector
for the $M$~direction and the $\mu$~direction, respectively.}
\begin{equation}
   S=\int\rmd^4x\,\left\{{1\over4}F_{MN}^aF_{MN}^a
   +{1\over2}\lambda^{aT}C\Gamma_MD_M\lambda^a\right\},
\label{twoxone}
\end{equation}
where
\begin{equation}
   F_{MN}^a
   =\partial_MA_N^a-\partial_NA_M^a+g_0f_{abc}A_M^bA_N^c,\qquad
   D_M\lambda^a=\partial_M\lambda^a+g_0f_{abc}A_M^b\lambda^c,
\label{twoxtwo}
\end{equation}
with $g_0$ being the dimensionless gauge coupling constant. In
eq.~(\ref{twoxone}), the matrix~$C$ is a charge conjugation matrix such
that\footnote{This is the matrix~$B_2$ of ref.~\cite{Inagaki:2004ar}.}
\begin{equation}
   C\Gamma_MC^{-1}=-\Gamma_M^T,\qquad
   C\Gamma C^{-1}=\Gamma^T,
   \qquad C^{-1}=C^\dagger,\qquad C^T=-C.
\label{twoxthree}
\end{equation}
Eq.~(\ref{twoxone}) is a Wick rotated version of the $\mathcal{N}=1$ SYM
theory in $d=4$ Minkowski spacetime. The above prescription for the Majorana
fermion in the euclidean space~\cite{Nicolai:1978vc,vanNieuwenhuizen:1996tv}
can be understood also from a view point of the ``Majorana decomposition''
which precisely gives rise to the ``half'' the Dirac fermion in the euclidean
space~\cite{Inagaki:2004ar}.

The action~(\ref{twoxone}) is invariant under the super transformation
\begin{eqnarray}
   &&\delta_\xi A_M^a=\lambda^{aT}C\Gamma_M\xi
   =-\xi^TC\Gamma_M\lambda^a,
\nonumber\\
   &&\delta_\xi\lambda^a={1\over2}F_{MN}^a\Sigma_{MN}\xi,
   \qquad\Sigma_{MN}={1\over2}[\Gamma_M,\Gamma_N],
\label{twoxfour}
\end{eqnarray}
owing to the Bianchi identity and the relation
$\Gamma_M\Gamma_N\Gamma_R=(1/3!)\Gamma_{[M}\Gamma_N\Gamma_{R]}
+\delta_{MN}\Gamma_R-\delta_{MR}\Gamma_N+\delta_{NR}\Gamma_M$. In addition
to this symmetry, eq.~(\ref{twoxone}) possesses the global $\U(1)_R$
symmetry
\begin{equation}
   \delta_\epsilon A_M^a=0,\qquad
   \delta_\epsilon\lambda^a=i\epsilon\Gamma\lambda^a,
\label{twoxfive}
\end{equation}
though this symmetry is broken by the anomaly.

By applying a dimensional reduction to eq.~(\ref{twoxone}), one can deduce
a euclidean version of the $\mathcal{N}=(2,2)$ SYM theory in 2~dimensions.
The dimensional reduction amounts to set $\partial_3\Rightarrow0$
and~$\partial_2\Rightarrow0$. To obtain a canonical normalization of fields,
we also rescale the gauge potentials and the gauge coupling as
$\ell A_M^a\Rightarrow A_M^a$ and~$g_0/\ell\Rightarrow g_0$, by using a
scale of length~$\ell$ which may be regarded as a size of the reduced (or
more appropriately compactified) directions. Then we regard $M=3$ and~$M=2$
components of gauge potentials as scalar fields as $A_3\Rightarrow\phi$
and~$A_2\Rightarrow\varphi$. The variable~$\lambda$ is mapped to the Dirac
fermion in 2~dimensions.\footnote{In a representation in $d=4$ in which
$\Gamma_0=\pmatrix{0&1\cr1&0\cr}$,
$\Gamma_i=\pmatrix{0&i\sigma^i\cr-i\sigma^i&0\cr}$,
$\Gamma=\pmatrix{-1&0\cr0&1\cr}$ and
$C=\pmatrix{i\sigma^2&0\cr0&-i\sigma^2\cr}$,
the Dirac field in $d=2$ can be defined from components of the
Majorana field by $\psi=\ell\pmatrix{\lambda_1\cr\lambda_2\cr}$ and
$\overline\psi=\ell(-\lambda_4,-\lambda_3)$, where
$\lambda=(\lambda_1,\lambda_2,\lambda_3,\lambda_4)$. The Dirac matrices
in $d=2$ then take the form, $\gamma_0=\sigma^3$, $\gamma_1=\sigma^2$ and
$\gamma=-\sigma^1$ and $B=\sigma^2$.}
After this dimensional reduction, fermion bi-linears are mapped to
\begin{eqnarray}
   &&\lambda^{aT}C\Gamma_\mu\mathcal{O}_{ab}\lambda^b
   \Rightarrow2\overline\psi^a\gamma_\mu\mathcal{O}_{ab}\psi^b\qquad
   \hbox{for $\mu=0$, 1},
\nonumber\\
   &&\lambda^{aT}C\Gamma_2\mathcal{O}_{ab}\lambda^b
   \Rightarrow2\overline\psi^a\gamma\mathcal{O}_{ab}\psi^b,
\nonumber\\
   &&\lambda^{aT}C\Gamma_3\mathcal{O}_{ab}\lambda^b
   \Rightarrow2\overline\psi^a(-i)\mathcal{O}_{ab}\psi^b,
\label{twoxsix}
\end{eqnarray}
where $\mathcal{O}$ is any anti-symmetric matrix with gauge and space
indices. In this way, we have a euclidean version of the $\mathcal{N}=(2,2)$
SYM theory in 2~dimensions:
\begin{eqnarray}
   S&=&\int\rmd^2x\,\biggl\{
   {1\over4}F_{\mu\nu}^aF_{\mu\nu}^a
   +{1\over2}D_\mu\phi^aD_\mu\phi^a
   +{1\over2}D_\mu\varphi^aD_\mu\varphi^a
   +{1\over2}g_0^2f_{abc}f_{ade}\varphi^b\phi^c\varphi^d\phi^e
\nonumber\\
   &&\qquad\qquad\mbox{}
   +\overline\psi^a\gamma_\mu D_\mu\psi^a
   -ig_0f_{abc}\overline\psi^a(\phi^b+i\gamma\varphi^b)\psi^c\biggr\}.
\label{twoxseven}
\end{eqnarray}
This action is invariant under the super transformation
\begin{eqnarray}
   &&\delta_\theta A_\mu^a
   =\overline\psi^a\gamma_\mu\theta-\overline\theta\gamma_\mu\psi^a,
\nonumber\\
   &&\delta_\theta\phi^a=-i\overline\psi^a\theta+i\overline\theta\psi^a,
   \qquad
   \delta_\theta\varphi^a=\overline\psi^a\gamma\theta
   -\overline\theta\gamma\psi^a,
\nonumber\\
   &&\delta_\theta\psi^a={1\over2}F_{\mu\nu}^a\sigma_{\mu\nu}\theta
   -i\gamma_\mu D_\mu(\phi^a+i\gamma\varphi^a)\theta
   +ig_0f_{abc}\varphi^b\phi^c\gamma\theta,
\nonumber\\
   &&\delta_\theta\overline\psi^a
   =-{1\over2}\overline\theta\sigma_{\mu\nu}F_{\mu\nu}^a
   -i\overline\theta\gamma_\mu D_\mu(\phi^a-i\gamma\varphi^a)
   +ig_0f_{abc}\overline\theta\gamma\varphi^b\phi^c,
\label{twoxeight}
\end{eqnarray}
(where $\sigma_{\mu\nu}=(1/2)[\gamma_\mu,\gamma_\nu]$; one notes that
$\sigma_{\mu\nu}=i\epsilon_{\mu\nu}\gamma$ with $\epsilon_{01}=1$ and
$\gamma\gamma_\mu=i\epsilon_{\mu\nu}\gamma_\nu$) which can be obtained by
applying the dimensional reduction to eq.~(\ref{twoxfour}). Note that in
eq.~(\ref{twoxseven}), $\psi^a$ and~$\overline\psi^a$ are regarded
independent variables. The $\U(1)_R$ symmetry~(\ref{twoxfive}) becomes the
fermion number symmetry
\begin{eqnarray}
   &&\delta_\epsilon A_\mu^a=0,\qquad
   \delta_\epsilon\phi^a=0,\qquad
   \delta_\epsilon\varphi^a=0,
\nonumber\\
   &&\delta_\epsilon\psi^a=-i\epsilon\psi^a,\qquad
   \delta_\epsilon\overline\psi^a=i\epsilon\overline\psi^a,
\label{twoxnine}
\end{eqnarray}
and, on the other hand, the rotational symmetry in the 2-3 plane becomes the
internal chiral symmetry
\begin{eqnarray}
   &&\delta_\epsilon A_\mu^a=0,\qquad
   \delta_\epsilon\phi^a=2\epsilon\varphi^a,\qquad
   \delta_\epsilon\varphi^a=-2\epsilon\phi^a,
\nonumber\\
   &&\delta_\epsilon\psi^a=i\epsilon\gamma\psi^a,\qquad
   \delta_\epsilon\overline\psi^a=i\epsilon\overline\psi^a\gamma,
\label{twoxten}
\end{eqnarray}
after the dimensional reduction.

\subsection{One-loop effective potential in the continuum theory}
As stated in Introduction, we correct a breaking of the supersymmetry in
our lattice formulation by supplementing scalar mass counter terms. To find
an appropriate value of a coefficient of counter terms, here we compute the
one-loop effective potential for scalar fields in the $\mathcal{N}=(2,2)$
SYM theory in 2~dimensions.\footnote{For this purpose, supersymmetric
Ward-Takahashi identities would be used as well. We found that, however, an
examination of the effective potential is much simpler.} For this
perturbative calculation, we add the gauge fixing term
\begin{equation}
   S_{\rm gf}=\int\rmd^2x\,
   {1\over2}\lambda_0\partial_\mu A_\mu^a\partial_\nu A_\nu^a,
\label{twoxeleven}
\end{equation}
to the action~(\ref{twoxseven}). With this gauge fixing term, the
Faddeev-Popov ghosts couple only to gauge potentials and the ghosts are
irrelevant to the present calculation of the one-loop effective potential
for scalar fields.

Perturbation theory in the present model, a massless theory in 2~dimensions,
is full of IR divergences and we need a careful treatment of zero modes.
For a reliable treatment of zero modes, we define the theory in a finite box
with size~$L$. We further impose the periodic boundary conditions for
all fields. The periodic boundary condition is consistent with the super
transformation~(\ref{twoxeight}) and invariance of the
action~(\ref{twoxseven}). Then the momentum becomes discrete and is given by
\begin{equation}
   p_\mu={2\pi\over L}n_\mu,\qquad n_\mu\in\mathbb{Z}.
\end{equation}
As usual, the one-loop effective potential for scalar fields is obtained by
performing gaussian integrations over fluctuations around the expectation
value of scalar fields. So we set
\begin{eqnarray}
   &&A_\mu^a(x)=\sum_p{1\over L}e^{ipx}\tilde A_\mu^a(p),
\nonumber\\
   &&\phi^a(x)=\phi^a+\sum_p{1\over L}e^{ipx}\tilde\phi^a(p),\qquad
   \varphi^a(x)=\varphi^a+\sum_p{1\over L}e^{ipx}\tilde\varphi^a(p),
\nonumber\\
   &&\psi^a(x)=\sum_p{1\over L}e^{ipx}\tilde\psi^a(p),\qquad
   \overline\psi^a(x)=\sum_p{1\over L}e^{ipx}\tilde{\overline\psi}^a(p),
\label{twoxthirteen}
\end{eqnarray}
where expectation values $\phi^a$ and~$\varphi^a$ are taken to be constant.
Substituting these into the action, (\ref{twoxseven})
plus~(\ref{twoxeleven}) and picking out terms quadratic in fluctuations, we
have
\begin{eqnarray}
   S+S_{\rm gf}
   &=&{1\over2}\sum_p\Bigl\{
   \tilde A_\mu(-p)[\delta_{\mu\nu}p^2-(1-\lambda_0)p_\mu p_\nu
   -\delta_{\mu\nu}(\Phi^2+\Psi^2)]\tilde A_\nu(p)
\nonumber\\
   &&\qquad
   \mbox{}+\tilde\phi(-p)(p^2-\Psi^2)\tilde\phi(p)
   +\tilde\varphi(-p)(p^2-\Phi^2)\tilde\varphi(p)
\nonumber\\
   &&\qquad
   \mbox{}+\tilde A_\mu(-p)ip_\mu\Phi\tilde\phi(p)
   +\tilde\phi(-p)ip_\mu\Phi\tilde A_\mu(p)
\nonumber\\
   &&\qquad
   \mbox{}+\tilde A_\mu(-p)ip_\mu\Psi\tilde\varphi(p)
   +\tilde\varphi(-p)ip_\mu\Psi\tilde A_\mu(p)
\nonumber\\
   &&\qquad
   \mbox{}+\tilde\phi(-p)(2\Psi\Phi-\Phi\Psi)\tilde\varphi(p)
   +\tilde\varphi(-p)(2\Phi\Psi-\Psi\Phi)\tilde\phi(p)
\nonumber\\
   &&\qquad
   \mbox{}+2\tilde{\overline\psi}(-p)
   (\gamma_\mu ip_\mu-i\Phi+\gamma\Psi)\tilde\psi(p)\Bigr\}+\cdots,
\label{twoxfourteen}
\end{eqnarray}
where we have introduced matrices
\begin{equation}
   (\Phi)_{ab}=g_0f_{acb}\phi^c,\qquad(\Psi)_{ab}=g_0f_{acb}\varphi^c,
\end{equation}
and abbreviated contractions in group indices. Gaussian integrations with
respect to fluctuations are straightforward and we begin with integrations
over zero-modes.

Gaussian integrations with respect to fluctuations with $p=0$ (zero modes)
give rise to the following contribution to the effective potential
\begin{eqnarray}
   &&{1\over L^2}\biggl\{\tr\log\{\Phi^2+\Psi^2\}
   +{1\over2}\tr\log\pmatrix{\Psi^2&-2\Psi\Phi+\Phi\Psi\cr
                             -2\Phi\Psi+\Psi\Phi&\Phi^2}
\nonumber\\
   &&\qquad
   \mbox{}-\tr\log\{\Phi^2+\Psi^2+i(\Psi\Phi-\Phi\Psi)\}
   \biggr\},
\label{twoxsixteen}
\end{eqnarray}
where $\tr$ denotes the trace with respect to group indices. In this
expression, the first line is a contribution from bosonic zero modes and the
second line comes from fermionic zero modes. One would expect that the above
three terms cancel out but this is not the case. This becomes clear by
considering configurations with $[\Psi,\Phi]=0$ or equivalently
$f_{abc}\varphi^b\phi^c=0$. These are nothing but configurations in the flat
directions along which the classical potential energy vanishes. For these
configurations, the first line of eq.~(\ref{twoxsixteen}) becomes singular
as~$\log0$ and the second line remains regular. Thus three terms in
eq.~(\ref{twoxsixteen}) do not cancel even at minima of the classical
potential.

The non-zero radiative corrections in the one-loop effective
potential~(\ref{twoxsixteen}) are {\it not\/} in contradiction with a
general property of supersymmetric theories that the vacuum energy vanishes
when the supersymmetry is not spontaneously broken.\footnote{In our present
problem, this property can be shown in the following way, for example: After
introducing the auxiliary field, the euclidean action can be expressed as a
super transformation of a certain gauge invariant operator. Assuming that
the supersymmetry is not spontaneously broken, this shows that the vacuum
energy is independent of the gauge coupling and may be adjusted to be zero.
The gauge fixing term and the Faddeev-Popov ghost term do not contribute to
the vacuum energy owing to the Slavnov-Taylor identity.} The effective
potential coincides with the vacuum energy only for minima of the effective
potential (because the external sources vanish only at minima of the
effective potential). Moreover, in our present case, at minima of the
classical potential, namely at configurations along the flat directions, the
quadratic term of bosonic zero modes acquires zero eigenfunctions and the
loop expansion (or the $\hbar$ expansion) breaks down.\footnote{It would
thus be inappropriate to call eq.~(\ref{twoxsixteen}) as the ``one loop''
effective potential for configurations along the flat directions.} To see
that the flat directions actually do not receive any radiative corrections,
one has to consider an integration over zero modes with a full part of the
action, not only the quadratic part.\footnote{This kind of study can be
found in ref.~\cite{Onogi:2005cz}.} In any case, irrespective of one's
interpretation on the ``one-loop'' radiative corrections from zero
modes~(\ref{twoxsixteen}), it is a very property of the target
{\it continuum\/} theory in a finite box and should be reproduced by any
sensible lattice formulation. We note that the
correction~(\ref{twoxsixteen}) vanishes and a naive expectation is
reproduced in the $L\to\infty$ limit.

We now turn to gaussian integrations over fluctuations with~$p\neq0$. For a
contribution of these non-zero modes, it is possible to
expand the effective potential with respect to expectations values. In the
quadratic order of $\phi^a$ and~$\varphi^a$ which will be relevant for later
discussions, we have
\begin{eqnarray}
   &&{1\over L^2}
   \biggl\{-\sum_{p\neq0}{1\over p^2}+\sum_{p\neq0}{1\over p^2}\biggr\}
   \tr\{\Phi^2+\Psi^2\}
\nonumber\\
   &&={1\over L^2}
   \biggl\{\sum_{p\neq0}{1\over p^2}-\sum_{p\neq0}{1\over p^2}\biggr\}
   g_0^2N_c(\phi^a\phi^a+\varphi^a\varphi^a),
\label{twoxseventeen}
\end{eqnarray}
where the first and the second terms in the parentheses come from bosonic
and fermionic fluctuations, respectively. Thus if we apply a uniform UV
regularization for bosonic modes and for fermionic modes, then a total
contribution to the effective potential vanishes. Obviously, this
cancellation is a result of an underlying supersymmetry. Note that this
result is independent of the gauge parameter~$\lambda_0$.

In summary, the one-loop effective potential in the $\mathcal{N}=(2,2)$ SYM
theory in 2~dimensions defined in a box of size~$L$ with the periodic
boundary conditions possesses the following properties. (1)~Contributions
from zero modes do not cancel out and take the form~(\ref{twoxsixteen}).
(2)~As eq.~(\ref{twoxseventeen}) shows, contributions from non-zero modes
to the quadratic parts precisely cancel out under a supersymmetric UV
regularization. In the next section, we determine a coefficient of mass
counter terms in our lattice formulation so that these properties of the
target theory are reproduced in the continuum limit.

\section{Lattice formulation of the $\mathcal{N}=(2,2)$ SYM theory in
2~dimensions}
\subsection{In the case of the Wilson fermion}
Our lattice action consists of three parts:
\begin{equation}
   S[U,\lambda]=S_{\rm G}[U]+S_{\rm F}[U,\lambda]+S_{\rm counter}[U].
\label{threexone}
\end{equation}
For bosonic fields, we use the standard plaquette action
\begin{eqnarray}
   &&S_{\rm G}[U]={1\over a^2g_0^2}\sum_{x\in\Gammait}\sum_{M,N}
   \Real\tr\bigl\{1-P(x,M,N)\bigr\},
\nonumber\\
   &&P(x,M,N)=U(x,M)U(x+a\hat M,N)U(x+a\hat N,M)^{-1}U(x,N)^{-1},
\label{threextwo}
\end{eqnarray}
where $U(x,M)\in\SU(N_c)$ represents the link variable. For the fermion
sector, we use the Wilson-Dirac operator~$D_{\rm w}$
\begin{equation}
   S_{\rm F}[U,\lambda]
   =-a^2\sum_{x\in\Gammait}
   \tr\{\lambda(x)CD_{\rm w}\lambda(x)\},\qquad
   D_{\rm w}={1\over2}\{\Gamma_M(\nabla_M^*+\nabla_M)
   -a\nabla_M^*\nabla_M\},
\label{threexthree}
\end{equation}
with covariant differences for the adjoint representation
\begin{eqnarray}
   &&\nabla_M\lambda(x)={1\over a}
   \left\{U(x,M)\lambda(x+a\hat M)U(x,M)^{-1}-\lambda(x)\right\},
\nonumber\\
   &&\nabla_M^*\lambda(x)={1\over a}
   \left\{\lambda(x)
   -U(x-a\hat M,M)^{-1}\lambda(x-a\hat M)U(x-a\hat M,M)\right\}.
\label{threexfour}
\end{eqnarray}
We use the overlap-Dirac operator in the next subsection. The last term in
eq.~(\ref{threexone}) is a mass counter term which will be specified below.

One verifies that by setting
\begin{equation}
   U(x,M)=\exp\{ag_0A_M^a(x)T^a\},
\label{threexfive}
\end{equation}
the classical continuum limit~$a\to0$ of eq.~(\ref{threexone}) without
$S_{\rm counter}$ is nothing but eq.~(\ref{twoxone}), the $\mathcal{N}=1$
SYM theory in $d=4$~dimensions, except overall powers of~$a$. In writing
eq.~(\ref{threexone}), we already performed the rescaling associated to the
dimensional reduction $d=4\to d=2$ by identifying $\ell=a$. So the gauge
coupling~$g_0$ in this section has a dimension of mass.

To realize a dimensional reduction from $d=4$ to $d=2$, we set the following
boundary conditions
\begin{eqnarray}
   &&U(x+a\hat M,N)=U(x,N),\qquad\lambda(x+a\hat M)=\lambda(x)\qquad
   \hbox{for $M=2$, 3},
\nonumber\\
   &&U(x+L\hat M,N)=U(x,N),\qquad\lambda(x+L\hat M)=\lambda(x)\qquad
   \hbox{for $M=0$, 1}.
\label{threexsix}
\end{eqnarray}
Namely, we reduce (or compactify) directions of $M=2$ and $M=3$ and impose
the periodic boundary conditions for other two directions. The size of the
2~dimensional box~$L$ is assumed to be an integer-multiple of the lattice
spacing~$a$. Thus our 2~dimensional lattice is given by
\begin{equation}
   \Gammait=\left\{x\in a\mathbb{Z}^2\mid 0\leq x_\mu<L\right\}.
\label{threexseven}
\end{equation}
The link variables are integrated with the invariant Haar measure
$\prod_{x\in\Gammait}\prod_M\rmd\mu(U(x,M))$ as usual.

Scalar fields in the $\mathcal{N}=(2,2)$ SYM theory in 2~dimensions are
identified with gauge potentials in $M=3$ and~$M=2$ directions:
\begin{equation}
   U(x,3)=\exp\{ag_0\phi^a(x)T^a\},\qquad
   U(x,2)=\exp\{ag_0\varphi^a(x)T^a\}.
\label{threexeight}
\end{equation}
The classical continuum limit of eq.~(\ref{threexone}) (without
$S_{\rm counter}$) with the boundary conditions~(\ref{threexsix}) reproduces
the $\mathcal{N}=(2,2)$ SYM theory in 2~dimensions~(\ref{twoxseven}). In our
lattice construction based on the dimensional reduction, scalar fields in
the $\mathcal{N}=(2,2)$ SYM theory in 2~dimensions are given by components
of link variables as eq.~(\ref{threexeight}). Therefore an integration along
these degrees of freedom is compact. In particular, the flat directions in
the target theory, along which the potential term
$(f_{abc}\varphi^b\phi^c)^2$ vanishes, become compact for finite lattice
spacings. Thus no subtlety is expected for numerical integrations along
these flat directions. A non-compactness of scalar fields in the target
theory is, as gauge potentials, restored in the continuum limit~$a\to0$.

Our proposal is similar to that of ref.~\cite{Maru:1997kh} at the point that
an extended SYM theory is formulated as a dimensional reduction of a lattice
formulation of the $\mathcal{N}=1$ SYM theory in a higher dimension.
Contrary to ref.~\cite{Maru:1997kh}, however, we do not claim no need of
tuning in a resulting low dimensional lattice theory. It is true that
$\mathcal{N}=1$ SYM theory in 4~dimensions~\cite{Curci:1986sm}, for example,
when formulated with the overlap-Dirac operator or with the domain-wall,
requires no fine tuning for a supersymmetric continuum
limit~\cite{Nishimura:1997vg}, owing to the exact lattice chiral symmetry.
After a dimensional reduction, however, a rotational symmetry among
reduced and un-reduced directions is violated and scalar mass terms are not
prohibited in general. In our formulation, this breaking of supersymmetry is
corrected by supplementing scalar mass terms.

In the continuum theory, a mixed mass term~$\varphi^a\phi^a$ is forbidden by
the chiral symmetry~(\ref{twoxten}) and only a symmetric mass term of the
form~$\phi^a\phi^a+\varphi^a\varphi^a$ is allowed (when the supersymmetry is
ignored). This persists in our lattice theory, owing to the exact discrete
symmetry
\begin{eqnarray}
   &&U(x,0)\to U(x,0),\quad U(x,1)\to U(x,1),\quad
   U(x,2)\to U(x,3)^{-1},\quad U(x,3)\to U(x,2),
\nonumber\\
   &&\lambda(x)\to\exp\left\{-{\pi\over4}\Sigma_{23}\right\}
   \lambda(x),
\label{threexnine}
\end{eqnarray}
which is a lattice analogue of the chiral rotation~(\ref{twoxten}) with
$\pi/4$ radian (recall that the chiral rotation was originally a space
rotation in the 2-3 plane). The scalar mass counter terms thus may be taken
as
\begin{equation}
   S_{\rm counter}[U]
   =-\mathcal{C}N_c\sum_{x\in\Gammait}\left(
   \tr\{U(x,3)+U(x,3)^{-1}-2\}
   +\tr\{U(x,2)+U(x,2)^{-1}-2\}\right),
\label{threexten}
\end{equation}
as this combination reduces to the symmetric mass term in the classical
continuum limit. Our task is therefore to determine an appropriate
coefficient~$\mathcal{C}$.

To determine~$\mathcal{C}$, we compute the one-loop effective potential of
scalar fields and compare it with that in the target theory in section~2.2.
For this perturbative calculation, we add the following gauge fixing
term\footnote{$\partial_\mu f(x)=(1/a)\{f(x+a\hat\mu)-f(x)\}$
and $\partial_\mu^*f(x)=(1/a)\{f(x)-f(x-a\hat\mu)\}$ denote the forward and
the backward differences, respectively.}
\begin{equation}
   S_{\rm gf}[U]=-a^2\sum_{x\in\Gammait}\sum_{\mu,\nu=0}^1\lambda_0
   \tr\{\partial_\mu^*A_\mu(x)\partial_\nu^*A_\nu(x)\},
\label{threexeleven}
\end{equation}
to the lattice action~(\ref{threexone}), where $\lambda_0$ being the gauge
parameter. As in the continuum theory, the ghost fields do not contribute to
the one-loop effective potential of scalar fields.

We also have to take account of the jacobian from the invariant group
measure $\prod_{x\in\Gammait}\*\prod_M\rmd\mu(U(x,M))$ for link variables to
a linear measure which is used in perturbation theory. That is given
by~\cite{Kawai:1980ja}
\begin{eqnarray}
   &&\prod_{x\in\Gammait}\prod_M\rmd\mu(U(x,M))
   =\prod_{x\in\Gammait}\prod_{M,a}\rmd A_M^a\,e^{-S_{\rm measure}},
\nonumber\\
   &&S_{\rm measure}=-\sum_{x\in\Gammait}\sum_M{1\over2}\tr\ln
   \left\{2{\cosh(a\mathcal{A}_M(x))-1
   \over a\mathcal{A}_M(x)}\right\},
\label{threextwelve}
\end{eqnarray}
where $\mathcal{A}_M$ is the gauge potential in the adjoint representation,
$(\mathcal{A}_M)_{ab}=g_0f_{acb}A_M^c$, and $\tr$ denotes the trace over
group indices. This factor gives rise to mass terms of scalar fields,
\begin{equation}
   S_{\rm measure}=a^2\sum_{x\in\Gammait}\left\{
   {1\over24}g_0^2N_c[\phi^a(x)\phi^a(x)+\varphi^a(x)\varphi^a(x)]+\cdots
   \right\},
\label{threexthirteen}
\end{equation}
that should also be included in the {\it one-loop\/} effective potential,
because this term is $O(g_0^2)$.

Now, to compute one-loop radiative corrections to the effective potential of
scalar fields, we substitute an expansion similar to
eq.~(\ref{twoxthirteen}) into the lattice action
$S_{\rm G}+S_{\rm F}+S_{\rm gf}$, with understandings (\ref{threexfive})
and~(\ref{threexeight}). In the present lattice case, the momentum~$p$ is
still discrete $p_\mu=(2\pi/L)n_\mu$ but is limited within the Brillouin zone
\begin{equation}
   \mathcal{B}=\left\{p\in\mathbb{R}^2\mid|p_\mu|\leq\pi/a\right\}.
\label{threexfourteen}
\end{equation}
We pick out terms quadratic in fluctuations and perform gaussian integrations
over fluctuations.

Let us first consider an integration over zero modes. A form of the lattice
action quadratic in zero modes is in general different from that in the
continuum~(\ref{twoxfourteen}) by terms of~$O(a)$. An integration over zero
modes in our lattice theory thus would give a different effective potential
from eq.~(\ref{twoxsixteen}). However, the difference is $O(a)$, because no
UV divergence arises from an integration over zero modes (these are finite
degrees of freedom). As a result, in the continuum limit $a\to0$, a
contribution of zero modes to the effective potential coincides with
eq.~(\ref{twoxsixteen}).

Next, we consider non-zero modes. The action quadratic in fluctuations is
\begin{eqnarray}
   &&S_{\rm G}+S_{\rm F}+S_{\rm gf}
\nonumber\\
   &&={1\over2}\sum_p\biggl\{
   \sum_{\mu,\nu}\tilde A_\mu(-p)
   \left[\delta_{\mu\nu}\hat p^2
   +(1-\lambda_0){1\over a^2}(e^{iap_\mu}-1)(1-e^{-iap_\nu})
   -\delta_{\mu\nu}\Phi^2\right]\tilde A_\nu(p)
\nonumber\\
   &&\qquad
   \mbox{}+\tilde\phi(-p)
   \left(\hat p^2+{1\over12}a^2\hat p^2\Phi^2\right)\tilde\phi(p)
   +\tilde\varphi(-p)(\hat p^2-\Phi^2)\tilde\varphi(p)
\nonumber\\
   &&\qquad
   \mbox{}+\sum_\mu\tilde A_\mu(-p){1\over a}(e^{ip_\mu a}-1)\Phi
   \tilde\phi(p)
   +\sum_\mu\tilde\phi(-p){1\over a}(1-e^{-ip_\mu a})\Phi\tilde A_\mu(p)
\nonumber\\
   &&\qquad
   \mbox{}+\tilde\lambda(-p)C
   (\tilde D_{\rm w}^{(0)}(p)+\Gamma_3\Phi-{1\over2}a\Phi^2)
   \tilde\lambda(p)\biggr\}+\cdots,
\label{threexfifteen}
\end{eqnarray}
where we have retained only terms relevant to a quadratic term in the
effective potential of~$\phi$. In the above expression,
$\tilde D_{\rm w}^{(0)}(p)$ is the momentum representation of the free
Wilson-Dirac operator
\begin{equation}
   \tilde D_{\rm w}^{(0)}(p)=i\Gamma_\mu\ring p_\mu+{1\over2}a\hat p^2,
\label{threexsixteen}
\end{equation}
and $\ring p_\mu$ and $\hat p_\mu$ denote momentum variables defined by
\begin{eqnarray}
   &&\ring p_\mu={1\over a}\sin(ap_\mu),\qquad
   \hat p_\mu={2\over a}\sin\left({1\over2}ap_\mu\right),
\nonumber\\
   &&\ring p^2=\sum_{\mu=0}^1\ring p_\mu^2,\qquad
   \hat p^2=\sum_{\mu=0}^1\hat p_\mu^2.
\label{threexseventeen}
\end{eqnarray}
By comparing eq.~(\ref{threexfifteen}) with eq.~(\ref{twoxfourteen}), we see
various form of lattice artifacts.

The gaussian integrations over non-zero mode fluctuations are
straightforward and, including a contribution of the measure
term~(\ref{threexthirteen}), we have
\begin{equation}
   {1\over L^2}\sum_{p\neq0}\left({1\over\hat p^2}
   -{1+{1\over2}a^2\hat p^2\over\ring p^2+{a^2\over4}(\hat p^2)^2}
   \right)g_0^2N_c\phi^a\phi^a
   \equiv-{1\over2}\mathcal{C}g_0^2N_c\phi^a\phi^a,
\label{threexeighteen}
\end{equation}
as the effective potential. The first term in the parentheses is a
contribution of bosonic fields and the second is a contribution of the
(Wilson) fermion. Note that this expression is independent of the gauge
parameter~$\lambda_0$. Comparing this with eq.~(\ref{twoxseventeen}), we see
that a cancellation of a bosons' contribution and a fermions' contribution is
not perfect owing to lattice artifacts. We correct this deviation from the
target continuum theory by adding the scalar mass counter
terms~(\ref{threexten}). An important observation is that
eq.~(\ref{threexeighteen}) is an IR as well as UV finite quantity. The
coefficient~$\mathcal{C}$ is a dimensionless number that depends only on the
ratio~$a/L$. In the continuum limit~$a\to0$, therefore, the summation can be
replaced by an integral
\begin{equation}
   {1\over L^2}\sum_{p\neq0}\to\int_{\mathcal{B}}{\rmd^2p\over(2\pi)^2},
\label{threexnineteen}
\end{equation}
and we have
\begin{equation}
   \mathcal{C}=-2\int_{-\pi}^\pi{\rmd^2p\over(2\pi)^2}\,
   \left({1\over\hat p^2}
    -{1+{1\over2}\hat p^2\over\ring p^2+{1\over4}(\hat p^2)^2}
   \right)=0.65948255(8)
\label{threextwenty}
\end{equation}
where we have rescaled the integration variables as $p_\mu\to p_\mu/a$ and
changed the definition of momentum variables as
$\ring p_\mu=\sin(p_\mu)$ and $\hat p_\mu=2\sin(p_\mu/2)$. This completes
our lattice formulation of the $\mathcal{N}=(2,2)$ SYM theory in
2~dimensions which uses the Wilson-Dirac operator. Namely, we claim that,
after including the mass counter terms~(\ref{threexten}) with the
coefficient~(\ref{threextwenty}), the target theory is obtained in the
continuum limit {\it without any further tuning of parameters}. We finally
remark that the lattice formulation in this subsection allows the strong
coupling expansion.

\subsection{In the case of the overlap fermion}
A use of the overlap fermion in our framework has a great practical
advantage because the fermion determinant is real and moreover
{\it semi-positive}. In this case, the fermion part of our lattice action is
given by
\begin{equation}
   S_{\rm F}[U,\lambda]
   =-a^2\sum_{x\in\Gammait}\tr\{\lambda(x)CD\lambda(x)\},
\label{threextwentyone}
\end{equation}
where the overlap-Dirac operator~$D$ is defined by
\begin{equation}
   D={1\over a}\{1-A(A^\dagger A)^{-1/2}\},\qquad
   A=1-aD_{\rm w},
\label{threextwentytwo}
\end{equation}
from the Wilson-Dirac operator~(\ref{threexthree}). As shown in
ref.~\cite{Neuberger:1997bg}, then the fermion determinant, or more
precisely the pfaffian, is semi-positive definite (see also
ref.~\cite{Kaplan:1999jn}; for a proof of this fact from general grounds,
see an appendix of ref.~\cite{Inagaki:2004ar}). Since we formulate the
$\mathcal{N}=(2,2)$ SYM theory in 2~dimensions by a dimensional reduction
of a 4~dimensional lattice gauge theory defined above, the fermion
determinant in 2~dimensional sense is also real and semi-positive even with
presence of scalar fields.

For the gauge sector, one may use the plaquette action~(\ref{threextwo}),
but from various point of view, the modified plaquette
action~\cite{Luscher:1998du,Fukaya:2003ph,Shcheredin:2004xa}
\begin{eqnarray}
   &&S_{\rm G}[U]={1\over a^2g_0^2}\sum_{x\in\Gammait}\sum_{M,N}
   \mathcal{L}(x,M,N),
\nonumber\\
   &&\mathcal{L}(x,M,N)=\cases{
   {\displaystyle\Real\tr\{1-P(x,M,N)\}\over
   \displaystyle1-\Real\tr\{1-P(x,M,N)\}/\epsilon'}
   &if $\Real\tr\{1-P(x,M,N)\}<\epsilon'$,\cr
   \infty&otherwise,}
\label{threextwentythree}
\end{eqnarray}
which dynamically imposes the
admissibility~\cite{Luscher:1981zq,Hernandez:1998et}, is more
appropriate.\footnote{In our present case of the adjoint fermion, the
overlap-Dirac operator is guaranteed to be well-defined if the admissibility
$\|1-\mathcal{P}(x,M,N)\|<\epsilon$ for all $x$, $M$ and~$N$, where
$\|A\|$ is the matrix norm, $\mathcal{P}^{ab}=-2\tr\{T^aPT^bP^{-1}\}$
denotes the plaquette in the adjoint representation and $\epsilon$ is a
certain constant, holds. Because of the inequality
$\|1-\mathcal{P}(x,M,N)\|\leq2\|1-P(x,M,N)\|\leq
2\sqrt{2\Real\tr\{1-P(x,M,N)\}}$, the action~(\ref{threextwentythree})
imposes this admissibility associated to the adjoint fermion by choosing
$\epsilon'\leq\epsilon^2/8$.}
Recall that, in our framework, the plaquette variable contains scalar fields
in the $\mathcal{N}=(2,2)$ SYM theory as well. Thus the admissibility
restricts also a configuration of scalar fields. One can confirm that,
however, this way of modification of the gauge action does not affect the
one-loop effective potential of the scalar field~$\phi$ in the continuum
limit (that we will compute). Thus for the gauge sector, we can use the
identical result as eq.~(\ref{threexeighteen}) (the first term).

For the fermion sector, we expand the action~$S_{\rm F}$ around the
expectation value to $O(\phi^2)$. Analogously to the last line of
eq.~(\ref{threexfifteen}), we have
\begin{eqnarray}
   S_{\rm F}&=&{1\over2}\sum_p\biggl\{\tilde\lambda(-p)C
   \biggl[\tilde D^{(0)}(p)+\tilde X(p)^{-1/2}(\Gamma_3\Phi-{1\over2}a\Phi^2)
\nonumber\\
   &&\qquad\qquad\qquad\qquad\mbox{}
   -{1\over4}a^3\tilde X(p)^{-3/2}\tilde A^{(0)}(p)\hat p^2\Phi^2\biggr]
   \tilde\lambda(p)\biggr\}+\cdots,
\label{threextwentyfour}
\end{eqnarray}
where
\begin{eqnarray}
   &&\tilde W(p)=1-{1\over2}a^2\hat p^2,
\nonumber\\
   &&\tilde A^{(0)}(p)=1-a\tilde D_{\rm w}^{(0)}(p)
   =\tilde W(p)-i\Gamma_\mu a\ring p_\mu,
\nonumber\\
   &&\tilde X(p)=\tilde A^{(0)}(p)^\dagger A^{(0)}(p)
   =\tilde W(p)^2+a^2\ring p^2,
\nonumber\\
   &&\tilde D^{(0)}(p)={1\over a}\{1-\tilde A^{(0)}(p)\tilde X(p)^{-1/2}\}.
\end{eqnarray}
As in the previous subsection, as a contribution of non-zero modes to the
one-loop effective potential, we have
\begin{equation}
   {1\over L^2}\sum_{p\neq0}\left\{{1\over\hat p^2}
   -{1\over
   \ring p^2+{1\over a^2}(\tilde X^{1/2}-\tilde W)^2}
   {\tilde W+a^2\ring p^2\over\tilde X^{1/2}}
   \right\}g_0^2N_c\phi^a\phi^a.
\label{threextwentysix}
\end{equation}
Thus, in the limit~$a\to0$, the coefficient of the mass counter
terms~(\ref{threexten}) is given by
\begin{eqnarray}
   \mathcal{C}&=&-2\int_{-\pi}^\pi{\rmd^2p\over(2\pi)^2}\,
   \left\{{1\over\hat p^2}
   -{1\over
   \ring p^2+(\tilde X^{1/2}-\tilde W)^2}
   {\tilde W+\ring p^2\over\tilde X^{1/2}}
   \right\}
\nonumber\\
   &=&-0.28891909(1)
\label{threextwentyseven}
\end{eqnarray}
where we have rescaled the integration variables as $p_\mu\to p_\mu/a$ and
changed the definition of momentum variables as
$\ring p_\mu=\sin(p_\mu)$, $\hat p_\mu=2\sin(p_\mu/2)$,
$\tilde W(p)=1-{1\over2}\hat p^2$ and~$\tilde X(p)=\tilde W(p)^2+\ring p^2$.
Thus, when the overlap-Dirac operator is used, a coefficient of the mass
counter terms~(\ref{threexten}) is given by eq.~(\ref{threextwentyseven}).

\subsection{UV power counting of lattice Feynman integrals}
In this subsection, we return to our argument based on the continuum power
counting in Introduction and confirm its validity. That is, with our lattice
action, we show that only one-loop scalar two-point functions are
potentially UV diverging. For this, we utilize the Reisz power counting
theorem~\cite{Reisz:1987da} on lattice Feynman integrals.

According to this theorem, the {\it superficial or overall\/} degree of UV
divergence of a lattice Feynman integral
\begin{equation}
   I_F=\int_{\mathcal B}\rmd^2k_1\cdots\rmd^2k_\ell\,
   {V(k,q;a)\over C(k,q;a)},
\end{equation}
where $q$ collectively denotes the external momenta, associated to an
$\ell$-loop Feynman graph~$F$, is given by
\begin{equation}
   \deg I_F=2\ell+\deg V-\deg C,
\end{equation}
where $\deg V$ and $\deg C$ are the UV degree of the numerator~$V$ and the
denominator~$C$, respectively. The UV degree for the numerator, $\deg V$, is
defined by the integer~$\nu$ in the asymptotic behavior
\begin{equation}
   V(\lambda k,q;a/\lambda)=K\lambda^\nu+O(\lambda^{\nu-1}),
\end{equation}
for $\lambda\to\infty$ and the UV degree of the denominator $\deg C$ is
similarly defined.

Now consider the Feynman rule resulting from our lattice action. From
expressions~(\ref{threexfifteen}) and~(\ref{threextwentyfour}), we
immediately find that propagators of bosonic fields (i.e., the gauge
potential and the scalar fields) contribute to $\deg C$ by~2 while the
fermion propagators (with either choice of Dirac operators) contribute to
$\deg C$ by~1. For interaction vertices, we find that purely bosonic
$b$~point vertices contribute to $\deg V$ by~$4-b$ at most and
interaction vertices of the fermion with $b$~bosonic lines $1-b$ at most.
From these, we have 
\begin{eqnarray}
   \deg I_F&\leq&
   2\ell+\sum_k n_k(4-b_k)+\sum_l\tilde n_l(1-\tilde b_l)
   -2i_b-i_f
\nonumber\\
   &=&2-{1\over2}i_f-\sum_kn_k(b_k-2)-\sum_l\tilde n_l\tilde b_l,
\end{eqnarray}
where we have assumed that the diagram~$F$ contains $n_k$ purely bosonic
vertices with $b_k$ boson lines and $\tilde n_l$ interaction vertices of the
fermion with $\tilde b_l$~bosonic lines. $i_b$ ($i_f$) denotes the number
of boson (fermion) internal lines and $e_b$ ($e_f$) denotes the number
of boson (fermion) external lines of the diagram. From the first line to the
second, we have used the ``topological relations''
$\ell=i_b+i_f-\{\sum_kn_k+\sum_l\tilde n_l-1\}$ and
$2\sum_l\tilde n_l=2i_f+e_f$.

By noting that $b_k\geq3$ and $\tilde b_l\geq1$, we see that the total number
of vertices, $\sum_kn_k+\sum_l\tilde n_l$, must be 1 or~2 for $\deg I_F$ to
be non-negative. Then it is easy to see that, besides vacuum bubbles, only
one-loop tadpoles (for which $I_F\leq1$) and two-point functions of bosonic
fields (for which $I_F\leq0$) may have a non-negative superficial degree of
UV divergence and thus potentially UV diverging. This conclusion is the same
as that from the continuum power counting. We then repeat the argument in
Introduction based on the gauge invariance and finally we infer that only
one-loop scalar two-point functions are potentially logarithmically UV
diverging and may suffer from lattice artifacts in the $a\to0$
limit.\footnote{We note that our power counting in this subsection which
concentrates only on UV divergences does not immediately lead to a rigorous
proof for our lattice action to possess a supersymmetric continuum limit to
all orders of perturbation theory. For such a proof, we have to properly
treat an effect of IR divergences associated to massless propagators in a
certain way. What we wanted to demonstrate here is that there do not emerge
``exotic'' lattice artifacts such as an example in Introduction which spoil
a conclusion of the continuum power counting.}

\subsection{Global symmetries}
Before concluding this section, we briefly comment on other global
symmetries besides the supersymmetry in our formulation. The target theory
possesses two $\U(1)$ symmetries, one is vector-like~(\ref{twoxnine}) and
another is chiral~(\ref{twoxten}). With a use of the Wilson-Dirac operator,
both symmetries are broken for finite lattice spacings. The fermion number
symmetry in 2~dimensions~(\ref{twoxnine}) was originally the {\it chiral\/}
$\U(1)_R$ symmetry in the $\mathcal{N}=1$ SYM theory in
4~dimensions~(\ref{twoxfive}) and this chiral symmetry is explicitly broken
by the Wilson term as usual. The chiral $\U(1)$ symmetry~(\ref{twoxten}), on
the other hand, was a rotation in the 2-3 plane in 4~dimensions and it is
broken by a lattice structure, though a discrete subgroup of it is preserved
by our construction as eq.~(\ref{threexnine}).

With a use of the overlap-Dirac operator, owing to the Ginsparg-Wilson
relation~$\Gamma D+D\Gamma=aD\Gamma D$~\cite{Ginsparg:1982bj}, the chiral
$\U(1)_R$ transformation in 4~dimensions~(\ref{twoxfive}) can be modified
as~\cite{Luscher:1998pq}
\begin{equation}
   \delta_\epsilon\lambda(x)
   =i\epsilon\Gamma\left(1-{1\over2}aD\right)\lambda(x),
\label{threextwentyeight}
\end{equation}
so that the lattice action is invariant even with finite lattice spacings.
Unfortunately, the fermion integration measure is not invariant under this
transformation, producing a non-trivial jacobian
\begin{equation}
   \exp\left\{
   a^2\sum_{x\in\Gammait}\epsilon{ia\over2}\tr\{\Gamma D(x,x)\}\right\}.
\label{threextwentynine}
\end{equation}
In the continuum limit, this jacobian becomes unity as is consistent with
the fact that the transformation~(\ref{twoxfive}) is
vector-like in a 2~dimensional sense (see eq.~(\ref{twoxnine})). For finite
lattice spacings, however, there is no reason to expect that the
jacobian~(\ref{threextwentynine}) is unity. As a result, the fermion number
$\U(1)$~(\ref{twoxnine}) as well as the chiral $\U(1)$~(\ref{twoxten}) is
not manifest in our formulation with the overlap-Dirac operator.

Note that, however, after adding the mass counter terms~(\ref{threexten}),
all correlation functions of elementary fields will coincide with continuum
ones in the continuum limit, as we have argued. Therefore, these $\U(1)$
symmetries will be restored in the continuum limit with {\it either\/} use
of Dirac operators without any further tuning of parameters.\footnote{In
correlation functions with some operators inserted, we have to examine
possible breaking of global symmetries caused by corresponding lattice
operators. We reserve this issue for future study.}

\section{Discussion}
In this paper, we proposed a lattice formulation of the $\mathcal{N}=(2,2)$
SYM theory in 2~dimensions, which appears to be favored from a view point of
numerical simulations. It must be possible to carry out Monte-Carlo
simulations with our formulation by present-day (or near-future) available
computer resources.\footnote{We have determined counter terms which ensure a
supersymmetric continuum limit. By a further systematic study along a
similar line, it might be possible to carry out the $O(a)$ improvement
program~\cite{Symanzik:1983dc} which accelerates an approach to a
supersymmetric continuum limit.} In fact, our construction starts with a
lattice formulation of the $\mathcal{N}=1$ SYM theory in
4~dimensions~\cite{Curci:1986sm}, which has intensively been studied by
numerical simulations~\cite{Montvay:2001aj}. For Monte-Carlo simulations to
be executable, we have to add a mass term for the fermion which explicitly
breaks the supersymmetry. The Majorana mass term in $d=4$, after the
dimensional reduction, becomes to the Majorana mass term for Dirac
fermions in $d=2$:\footnote{In this expression, $B$ denotes the
``charge conjugation matrix'' in $d=2$ such that
$B\gamma_\mu B^{-1}=-\gamma_\mu^T$, $B\gamma B^{-1}=-\gamma^T$,
$B^{-1}=B^\dagger$ and $B^T=-B$; this is the matrix~$B_1$ of
ref.~\cite{Inagaki:2004ar}.}
\begin{eqnarray}
   S_{\rm mass}[\lambda]
   &=&-a^2\sum_{x\in\Gammait}im\tr\{\lambda(x)C\Gamma\lambda(x)\}
\nonumber\\
   &\Rightarrow&
   a^2\sum_{x\in\Gammait}\left(
   -m\tr\{\psi(x)B\psi(x)\}
   +m\tr\{\overline\psi(x)B\overline\psi(x)\}\right).
\label{fourxone}
\end{eqnarray}
(With the overlap-Dirac operator, the fermion determinant is positive
definite with this mass term~\cite{Inagaki:2004ar}.) Thus, we have to take
the massless limit~$m\to0$ in addition to the continuum limit~$a\to0$ in
simulations.

What kind of observable will be interesting to be explored in numerical
simulations? An obvious candidate is mass spectrum of bound states. Mass
spectrum and the two-point function of the energy momentum tensor in this
$\mathcal{N}=(2,2)$ SYM theory have been
investigated~\cite{Antonuccio:1998mq} by using the supersymmetric discrete
light-cone quantization~\cite{Matsumura:1995kw}; there, a closing of the
mass gap, in accord with an argument~\cite{Witten:1995im} on the basis of
the 't~Hooft anomaly matching condition, is reported. Numerical simulations
based on a lattice formulation should be confronted with these results.
Before going to study these physical observables, of course, we should be
sure about our argument on a restoration of supersymmetry. Thus a
restoration of supersymmetric Ward-Takahashi identities has to be firstly
confirmed.

Clearly, one is interested in a generalization of our present proposal to
other low dimensional extended SYM theories which have been formulated in
refs.~\cite{Kaplan:2002wv,Sugino:2003yb,Catterall:2004np}; the
$\mathcal{N}=(4,4)$ and $\mathcal{N}=(8,8)$ SYM theories in 2~dimensions and
the $\mathcal{N}=2$, $\mathcal{N}=4$ and~$\mathcal{N}=8$ SYM theories in
3~dimensions, all these can be obtained by the $\mathcal{N}=1$ SYM theory in
higher dimensions via the dimensional reduction. For this issue, it is useful
to distinguish two aspects of our present formulation. First, we have
defined the $\mathcal{N}=(2,2)$ SYM theory in 2~dimensions by using a
dimensional reduction of the $\mathcal{N}=1$ SYM theory in 4~dimensions. In
a similar way, the $\mathcal{N}=2$ SYM theory in 3~dimensions may be
formulated starting with eq.~(\ref{threexone}) just by reducing only $M=3$
direction, although for this case, we need more general counter terms. (A
coefficient of these counter terms is expected to be UV finite, from a power
counting and the gauge invariance.) A generalization of this aspect of our
formulation to the $\mathcal{N}=1$ SYM theory in 6~dimensions and in
10~dimensions is however not straightforward. For the $\mathcal{N}=1$ SYM
theory in 6~dimensions, we first have to define a lattice gauge theory in
6~dimensions which contains a single adjoint Weyl fermion. This theory,
in a 6~dimensional gauge theoretical sense, is anomalous and we first have
to show that possible obstructions in a gauge invariant lattice formulation
of chiral gauge theories, such as the one in
refs.~\cite{Luscher:1998du,Luscher:1999un}, disappear in a process of the
dimensional reduction $6\to2$ or $6\to3$. (The gauge anomaly in general
implies a failure~\cite{Matsui:2004dc} of a lattice formulation along lines
of refs.~\cite{Luscher:1998du,Luscher:1999un}.) It is conceivable that this
can be shown by imitating an argument of~ref.~\cite{Suzuki:2000ku}. A
generalization to the $\mathcal{N}=1$ SYM theory in 10~dimensions seems much
harder because we do not have a proper local lattice action for the
Majorana-Weyl fermion in 10~dimensions for the
present~\cite{Inagaki:2004ar}. 

On the other hand, we used the fact that the $\mathcal{N}=(2,2)$ SYM theory
in 2~dimensions is super-renormalizable and argued that one-loop calculable
mass counter terms ensure a supersymmetric continuum limit. The above SYM
theories in 2 and~3 dimensions are all super-renormalizable and thus our
argument is equally applied to the above list of theories, although
3~dimensional theories require various type of counter terms besides scalar
mass terms. Reality and positivity of the fermion determinant and a
compactness of flat directions are of course a separate issue and we have to
find some mechanism if these properties desirable for numerical simulations
are thought to be kept.

\paragraph{Note added.}
Quite recently, Elliott and Moore~\cite{Elliott:2005bd} performed
independently a similar analysis to ours for the $\mathcal{N}=2$ Wess-Zumino
model and the $\mathcal{N}=2$ supersymmetric QCD in 3~dimensions.

\acknowledgments
We would like to thank Takanori Fujiwara, Nobuyuki Ishibashi, Koichi
Murakami, Tadakatsu Sakai, Fumihiko Sugino and Akira Ukawa for helpful
discussions. This work is supported in part by Grant-in-Aid for Scientific
Research, \#13135203.


\listoftables           
\listoffigures          

\end{document}